\documentclass[creativecommons]{eptcs}
\providecommand{\event}{UITP 2014}

\usepackage[english]{babel}
\usepackage{listings}
\usepackage{graphicx}

\title{PIDE for Asynchronous Interaction with Coq}
\def\titlerunning{PIDE for Coq}
\author{Carst Tankink\thanks{Supported by the Paral-ITP project: ANR-11-INSE-001 \url{http://paral-itp.lri.fr/}}
\institute{INRIA Saclay - \^Ile de France}
\email{carst.tankink@inria.fr}
}
\def\authorrunning{Carst Tankink}

\begin{document}
\lstset{language=XML,basicstyle=\small\ttfamily,escapechar=!}
\maketitle

\begin{abstract}
  This paper describes the initial progress towards integrating the Coq proof assistant with the PIDE architecture initially developed for Isabelle~\cite{Makarius-2014}. The architecture is aimed at asynchronous, parallel interaction with proof assistants, and is tied in heavily with a plugin that allows the jEdit editor to work with Isabelle.

  We have made some generalizations to the PIDE architecture to accommodate for more provers than just Isabelle, and adapted Coq to understand the core protocol: this delivered a working system in about two man-months.
\end{abstract}

\section{Introduction}
Since October 2008, the Isabelle proof assistant has an architecture aimed at asynchronous proof processing.
% TODO: find citation to support this
To support this architecture, the tool also gained a significant update to its user interface in the form of Isabelle/jEdit: a plugin for jEdit that supports interaction with Isabelle.
jEdit is a generic editor aimed at programmers in any language: the editor's main components are a group of text windows with syntax highlighting, and any other functionality (compilation, version control, project management, \ldots) can be added through the use of plugins.
Isabelle's support for jEdit was added through this plugin system, by adding a `generic' prover plugin that exchanges messages with the proof assistant in order to provide a number of functions for authoring proofs, such as:

\begin{enumerate}
  \item prover output;
  \item syntax highlighting;
  \item `semantic' highlighting, provided by the prover as it processes the text. A prominent example is the addition of links, which point to the location where an entity (such as a lemma) is defined. These links are attached to the locations where this definition is used.
\end{enumerate}

\begin{figure}
  \includegraphics[width=\textwidth]{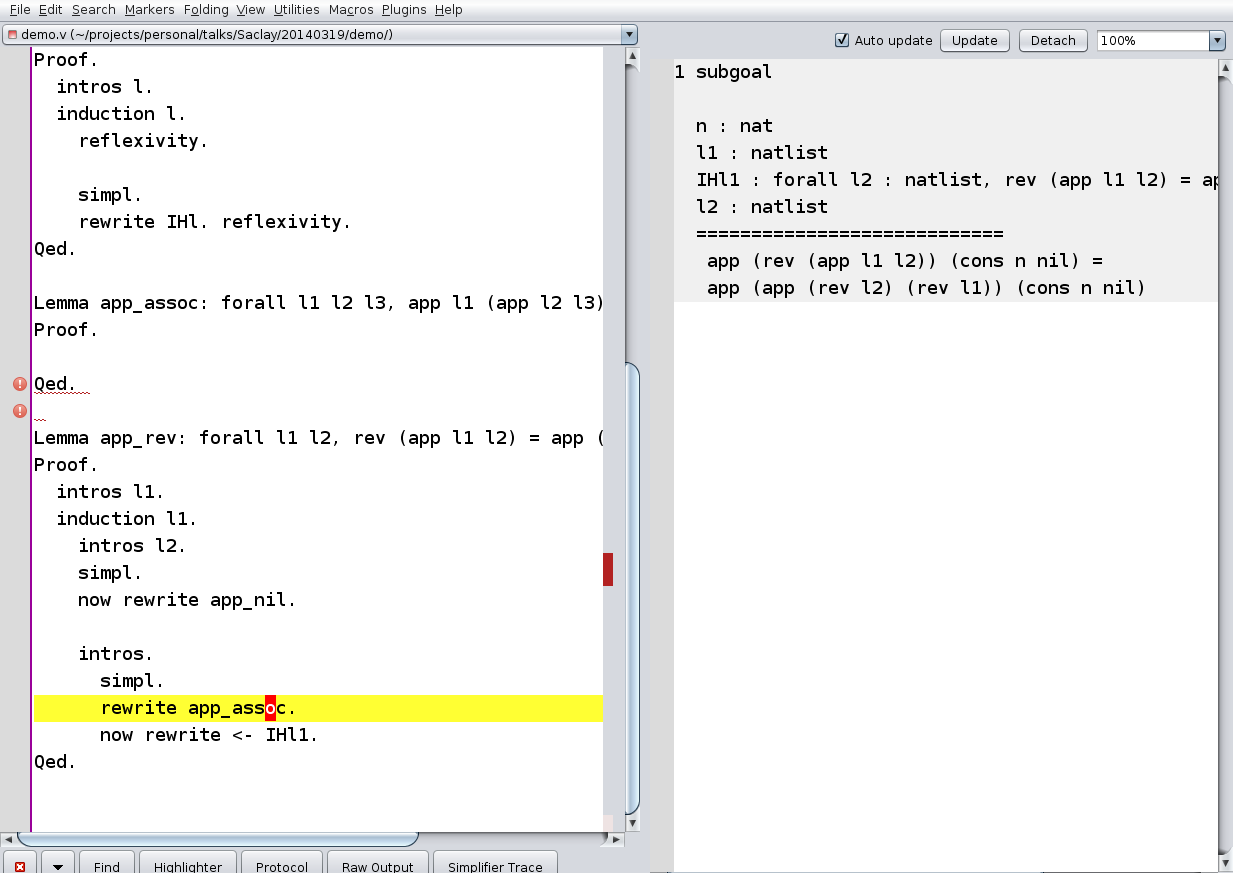}
  \caption{The jEdit plugin working with Coq}
  \label{fig:jEdit-Coq}
\end{figure}

% TODO: Citations/URLs of previous experiments.
While the plugin aims to be generic, this claim has never been validated in anger: the main development has centered on Isabelle, and while there have been experiments with Coq~\cite{Makarius-2013} and Why3, these did not invoke actual prover processes, but only provided a shallow interface to (stateless) syntax highlighting.
This paper describes the effort, started in February 2014, to integrate Coq with this plugin, by adapting to the generic PIDE architecture as much as possible, and driving further generalizations in that architecture where necessary. The resulting interface is shown in Figure~\ref{fig:jEdit-Coq}. This paper also shows a relatively outsider view of the PIDE architecture, and might therefore be useful to others interested in integrating their formal tools with the jEdit prover plugin.

\section{PIDE and Isabelle/jEdit}
% Extended intro: show the architecture we are working with
To bridge jEdit, written in Java, with Isabelle, written in Standard ML, Makarius Wenzel developed a pair of libraries and an underlying protocol called PIDE: it exposes a set of functions in either language, and these functions are translated into an internal protocol, used to transfer data between the specific programs. The protocol is used to propagate text edits from the editor to the proof assistant, and to send feedback to the editor, coming from the proof assistant.

The basic design of PIDE is as follows, looking from the JVM side (the ML view is symmetrical).
\begin{enumerate}
  \item At the highest level, there are a number of functions, capturing a \emph{Document}: such a document represents all versions of a proof text, and can be updated by using these functions. Additional functions are `hooks' that allow the tool to receive changes in the document made by the prover.
  \item These functions, and the data taken as arguments and results, are represented as XML, that can be serialized and de-serialized by both sides.
%TODO: Citation for Yxml.
  \item The XML encoding is translated to YXML, a custom transfer syntax for XML.
  \item Finally, the YXML packets are sent on a pair of input/output channels. In Unix-like systems, these are FIFO buffers, but the PIDE implementation also allows for communicating over TCP/IP.
\end{enumerate}

PIDE can been used to build any (JVM-based) tool that uses the proof assistant as some black box, but the flagship application is the Isabelle plugin for jEdit. Building on top of this library, this plugin provides functionality for interacting with the prover from jEdit: it listens to changes in the text areas of the editor, parses these changes into `command spans', and uses the PIDE functions to send the spans onwards. It uses the prover hooks to react to the proof being processed, storing prover states to be rendered when the user focuses on the corresponding command span, and using `semantical' information to add highlighting and hyperlinks to the text. The functionality of the plugin goes beyond this, but this is the core functionality we are interested in bringing to Coq.

\section{Generalizing PIDE}
% Reusable components of PIDE:
% - Parser architecture: new command parser for Coq.
% - All markup: requires adapting Coq to send appropriate messages
% - jEdit integration, obviously.plus_S_r thm
We want to use Coq with the architecture, obtaining the interaction model with as little adaption as possible: we want to reuse building blocks from the existing PIDE and plugin implementations, by generalizing code where possible and writing Coq-specific code where necessary.
With that in mind, we have a choice of where exactly to insert the desired functionality. In theory, it is enough to just implement the high-level interface with Coq-specific protocols. Such an interface is required to send textual changes to the prover.
It is also required to provide a `snapshot' of the proof state: this snapshot is iterated over by the editor side of the plugin, and is used to determine what information the prover has on certain spans.
This information is then used to provide the various displays of the proof document.

Implementing these functions from scratch is certainly possible, but is a large effort leading to a partial duplicate of the PIDE implementation for Isabelle.
Instead, we take a better look at the elements of the PIDE architecture, focussing on how individual parts can be reused.

\subsection{Change parsing}
The change parsing `algorithm' takes a sequence of text edits (insertions and deletions) and uses these to determine what the new command spans for the prover are. To do so, the algorithm takes several steps, when a non-empty list of edits is provided. The following description is based on an analysis of the file \texttt{src/Pure/Thy/thy\_syntax.scala} of the Isabelle source code, changeset \texttt{bbf4d51}. This source file is available through \url{https://bitbucket.org/Carst/pide_notes/src/bbf4d51/src/Pure/Thy/thy_syntax.scala}.

The algorithm works on a previous version of the document, and creates a new version by applying the edits in a specific manner. The steps taken on a specific document are described below.

\begin{enumerate}
  \item Determine any changes to the header of the proof document. This header is an Isabelle-specific syntax element, that declares the name of the proof document and the other documents it imports. Determining the changes here is important there, because a dependency can define new commands, that direct the parsing in the next step.
  \item The second step `applies' the edits to the document, as well as reparsing any text that might have been affected by the header changes. We only focus on the first part here. In this procedure, the command spans that are directly affected by an edit are manipulated to reflect the new text, and following this, reparses the resulting text in order to obtain the new command spans.
\end{enumerate}

The first step is specific to Isabelle, but a large part of the second step is generic: it creates a new document version, and runs a language-specific parser in order to obtain the command spans.  This parser is a part of the Scala side of PIDE and is used to make certain computations quicker, for example syntax highlighting. It also reduces the communication overhead between the Java-based processes and the prover-specific ones: all communication can be done in terms of pre-defined command spans. This means that we can run this algorithm from our Coq implementation of PIDE, but provide a Coq-specific parser that translates text into command spans. The resulting parser is straightforward: it tokenizes the input, looking for the Coq-specific terminator symbol, the period (`.'). Its implementation can be found at \url{http://tinyurl.com/coq-loader}\footnote{These tinyurls point to specific source files in the bitbucket repository \url{https://bitbucket.org/Carst/pide_notes/src//src/Pure/Coq/}.} (the high-level parser interface) and at \url{http://tinyurl.com/coq-parser} (the actual tokenizer). To allow using this tokenizer with the original change parser, the algorithm had to be changed, to better display this structure: the change parsing step would call the Isabelle parser directly, but after a change, made in collaboration with Wenzel, the algorithm can call any prover-specific parser that implements a generic interface.

The benefit of using this algorithm, is that the data structure that is used to store the results is directly usable by the standard protocol of the PIDE implementation: the resulting changes on the prover level are transformed all the way down to the YXML syntax, and sent to the prover.
We did need to implement the protocol stack on the Coq side, which is described in Section~\ref{sec:AdaptingCoq}.

\subsection{Markup accumulation}
The accumulation of prover feedback is done by a generic function that collects all information in the document in an data structure. This data structure is then queried by the plugin to generate markup.
The accumulated feedback is expected in a format that was designed for the Isabelle implementation, but can still be mimicked by Coq because most of the specifics are in naming conventions within the messages: these names are, for example, tags that provide markup to a certain region of text, using nomenclature that is standard in the Isabelle ecosystem, but not that of Coq.

\section{Adapting Coq}
\label{sec:AdaptingCoq}

To make Coq compatible with the PIDE design, we need to add a number of features to the `top level' of the prover.
Until now, all interaction with Coq has been in a procedural `Read-Eval-Print-Loop' (REPL): any interface to the system
was required to provide both the content of the proof as well as the commands that instruct the prover to process this content.
These two types of commands have always been unified in a single transaction: the interface would send a single command span to the prover,
the system would process the span, and after that return control to the interface, which could then send a new span.

From an interaction perspective, this is far from ideal, since the user will have to undo steps in order to add information to a proof, and this information might not impact the rest of the proof directly.
From a prover perspective, this state of affairs is also not ideal: when the system `discovers' a proof line by line, it cannot schedule the processing of the lines in any meaningful way. In particular, in the face of parallel computation, it is impossible for the system to pass off computation of an expensive line in order to start working on a next, unrelated line of the proof.

\begin{figure}
  \includegraphics[width=\textwidth]{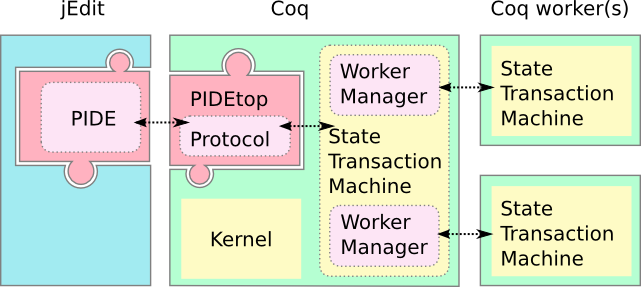}
  \caption{Architecture of PIDE and Coq \label{fig:arch}}
\end{figure}

To accomodate for the asynchronous interaction model of PIDE, the toplevel
therefore needs to implement a protocol that takes changes to the idealized
document and propagates these to a computation engine (the State Transaction
Machine described in Section~\ref{sec:STM}). This state of afairs is displayed
in Figure \ref{fig:arch}. The architecture shows the PIDE plugin for jEdit, and
the adapted toplevel and the State Transaction Machine with its worker processes.

\subsection{State Transaction Machine}
\label{sec:STM}

In order to make the system more suitable for PIDE, Enrico Tassi previously~\cite{Tassi-2012} developed a ``State Transaction Machine'' (STM) for Coq. This machine no longer depends on a linear, line-by-line model of a proof, but rather provides a computation model based on a DAG: the interface still adds commands to this DAG, but these commands are parsed into a graph that separates commands that can be executed out of order. The commands do not automatically trigger a computation, but the interface can request to start the computation by ``observing'' any node in the graph. The system will then compute all dependencies of this node, as well as the node itself.
The computation itself can now be scheduled by the STM based on this graph, and the computation is scheduled to be executed by slave processes running in parallel. As a result, the computed proof states are also reported asynchronously.

\subsection{PIDE for Coq}
While the PIDE protocol for Isabelle, which we use here, sends out several different types of messages, it is only necessary to implement a reaction to two messages in order to obtain a working integration with the jEdit interface. Symmetrically, the prover needs to send two messages in order to obtain any interaction with the user.

The protocol stack is modelled similarly to the implementation in Isabelle: its lowest layer is a raw FIFO-channel, the messages are parsed into YXML, which is translated to XML. Finally, XML messages are interpreted and mapped to function calls intended for Coq. This architecture is based on the prototype implementation described before by Wenzel~\cite{Makarius-2013}, but is extended with hooks that trigger the prover's computation. Using this organization gave us a running start, as the parts that deal with maintaining an internal `PIDE document' are already in place.

Currently, we do  not mimic the syntax highlighting described in the original architecture. Its implementation was based on the ad-hoc syntax highlighter used for the CoqIDE interface for Coq, and introduced an unwanted dependency on the GTK graphical library. The library is unwanted because it is not used to provide any graphics, and is an optional component to the Coq installation. While it is possible to eliminate the dependency, the resulting highlighting would be rather limited. As a result, it seems better to reconsider the syntax highlighting completely, and integrate it more deeply in the implementation of Coq itself, which would have the added benefit of allowing other interfaces for Coq to use the information.
An initial attempt is currently in progress that makes the syntax tree available to the PIDE protocol implementation, which allows highlighting the syntax robustly. These efforts are still in a too early stage to be able to report any results of this approach, either positive or negative.

The adaptations made to Coq to accommodate this protocol can be found in the following branch of the Coq source tree: \url{https://bitbucket.org/Carst/coq-pide/src/?at=PIDE}. The files of interest are all contained in the ``pide'' directory. Of particular interest are the files \href{https://bitbucket.org/Carst/coq-pide/src/b3f587d40989b0dfa9962e2f94d96ff98115088a/toplevel/pide_protocol.ml?at=PIDE}{\texttt{pide\_protocol.ml}} and \href{https://bitbucket.org/Carst/coq-pide/src/b3f587d40989b0dfa9962e2f94d96ff98115088a/toplevel/pide_document.ml?at=PIDE}{\texttt{pide\_document.ml}}, which implement the high-level protocol.

The Coq implementation follows the Isabelle version of PIDE loosely: it maintains a version history of Documents. Each document version represents the state of the proof at a particular moment in time, both its content and the computations executed on them. For a more complete description of this architecture, we refer to Makarius Wenzel's paper on PIDE~\cite{Makarius-2014}.

\paragraph{Define commands}
The first type of message coming from the interface maps identifiers to the contents of command spans. All future messages from the interface do not refer to the text, but instead refer to the command identifiers.

To handle this situation, Coq maintains a lookup table for the defined commands in the document module, that allows it to search for command spans whenever an identifier is given.

\paragraph{Update}
The update message is the center of the protocol: it notifies the prover of the changes by giving the insertions and deletions as command identifiers. In Coq, we use this message to update an internal representation of the proof document, by executing the insertions and deletions of the edit. Afterwards, we map each command in the new document to a future computation of a proof state, retaining the computations that have been executed on the prefix of commands that are unchanged between the two versions: currently, we do not attempt to keep computations on shared commands after the first difference, because the state of these computations can be changed by the differences. At this stage, we only prepare the identifier for that computation, and return this in an `assign update' message: this is the only message expected by the interface: it confirms that the prover received the update, and notifies the interface of which computations will take place: future messages from the computation will be tagged with this execution identifier.

After the assignment is reported to the interface, Coq continues by adding the new commands to the STM. Once all commands have been executed, it `observes' the final command of the proof. This triggers the computation of the proof states.

\paragraph{Reporting States}
When Coq is computing, the states are reported asynchronously, together with the execution identifier and the offset of the part of the command that was computed to obtain this state. The system then wraps this information in a `writeln' message, and sends it on to the Scala side. The PIDE plugin on the jEdit side is equipped to handle these messages, and attach them to the correct piece of text in the editor: when the user places the text cursor on a span, the corresponding state is shown.

Should an error occur while processing the command, this is reported asynchronously, but now wrapped in an `error' message.
These are reported differently in the jEdit interface, by placing a `squiggly underline' (similar to a spell-checker's in a text editing program) on the offending region: this is shown in Figure~\ref{fig:jEdit-Coq}.
Also note that this truly is an asynchronous report: despite an error occurring on the `Qed' line of the lemma ``app\_assoc'', the user is still able to inspect the state after the highlighted line, which \emph{uses} the `broken' lemma: its statement is still available at a later point in the proof.

Another application of asynchronous printing is the ability to report more than just the state after executing a tactic. Coq allows the user to compose proof steps
into larger tactics that work together. A simple example of this is sequential composition, which runs each proof step on the result of the previous proof step.
In the REPL model, executing such a composed tactic would only show the final state, which is not useful when the reader of a proof script wants to understand
how the composed tactic works. As an alternative, the editor could evaluate such a tactic in small steps, but this reduces the effectiveness of combining tactics in
the first place. PIDE's asynchronous reporting mechanism provides an elegant solution to this conundrum: composed tactics can report the intermediate results from
their components, which can then later be inspected by the user.

\paragraph{Markup}
Finally, the PIDE protocol allows, but does not require, reporting markup. The details of the implementation are very prover specific, and we will not go too deeply into it here. What is worth mentioning, is that we could use the messaging system to obtain hyperlinks for the Coq sources: in the jEdit interface, it is possible to click on a name of a lemma (or theorem, or definition, or \ldots), and be taken to the position where this entity is defined.
This was achieved by exporting information in Coq, using the `report' messages that are part of the PIDE protocol for Isabelle: the plugin is equipped to handle these messages, and transform them into source code hyperlinks.

% Coq has a more document-oriented/declarative architecture since 2012 (Tassi).
% Changes in the prover become changes to the maintained document
% - Currently use the actual PIDE messages, interpreted for Coq.
% - Translation of Coq internals: allows hyperlinking, keywords (maybe?)
% - Use of asynchronous message-passing to report to editor.

\section{PIDE in action}
To give a better account of the PIDE and STM architecture, we give an overview of the messages and function invocations used to obtain Figure~\ref{fig:jEdit-Coq}. We do truncate messages occasionally, and do not use the actual identifiers, as they change per session. The courageous reader can see all the protocol messages in action by installing and running the jEdit plugin with Coq, while having the 'Protocol' window open.

\paragraph{Defining commands} Modifying text (for example, by keyboard entry, copy-paste or by loading a file) triggers an update event from the buffer to the PIDE plugin. This event specifies what region has changed. PIDE then transforms its internal representation based on this event, by removing and inserting text where necessary and then (re)parsing the text affected by the change: at least inserted text needs to be parsed, but a larger region might need to be modified, for example when inserting or removing an `open comment' mark. The language-specific parser returns command spans: for Coq these are all phrases ending with a single period, including the following whitespace\footnote{For brevity's sake we do not include whitespace in the examples} and preceding comments: ``Proof.'', ``intros l.'', \ldots.

The interface then sends the commands to the prover, each in a separate XML message (the empty arguments are used by Isabelle):
\begin{lstlisting}
<prover_command name="Document.define_command">
<prover_arg>-1</prover_arg>
<prover_arg/>
<prover_arg/>
<prover_arg>Proof.
  </prover_arg>
</prover_command>
\end{lstlisting}

This message triggers a function in the prover that stores the key-value pair (``-1'', ``Proof.'') into the lookup table for commands. The key, like any identifier coming from the prover, is a negative number: the prover uses positive numbers to report its identifiers.

\paragraph{Updating document state} After all define\_command messages have been sent out, the prover sends a ``Document.update'' message (heavily truncated):
\begin{lstlisting}
<prover_command name="Document.update">
<prover_arg>0</prover_arg>
<prover_arg>-1492</prover_arg>
<prover_arg>
  <:>
    !\vdots!
    <:>/home/carst/projects/pide/tests/foo.v</:>
    <:><0>
        <:>
          <:/>
          <:><:>-1</:></:>
        </:>
        <:>
          <:><:>-1</:></:>
          <:><:>-2</:></:>
        </:>
        !\vdots!
    </0></:>
    !\vdots!
  </:>
</prover_arg>
</prover_command>
\end{lstlisting}

This message triggers a function with three arguments. The first is the version of the document to update from, the second is the version to upgrade to.
The third argument is a list of update operations (the `\lstinline!<:>!' tags are used as transfer separators for list elements and pairs), consisting of a file name to apply the operation on, and the operation itself: there are several types of operations in PIDE, but the only one implemented for Coq (and the only one necessary for a functioning PIDE-compatible prover) is operation type `0': the Edit operation.  This edit operation consists of a list of pairs of optional command identifiers: the first operation encodes the pair (None, -1), which means inserting the command with identifier `-1' at the beginning of the document. The second pair is (-1, -2), which encodes an insertion of the command identified by -2 after the command with identifier -1. The third type of pair, not displayed in this message, would have an identifier as its first element, and `None' as its second: this encodes the deletion of the command directly after the command identified by the first element.

The triggered update function looks up the old version of the proof `document' (a list of pairs of command identifiers and execution identifiers) and inserts the new version of the document: it progressively applies the insertions and deletions encoded in the message on the old version. In this case, the old version is empty, and the function starts with the insertion of command `-1'. It then continues on the resulting document, inserting the command `-2' after the `-1' just inserted, obtaining the sequence `-1; -2'\footnote{In functional programming terms, update is a fold over the list of update operations, starting at the old version of the document}. After obtaining the new document version, the function assigns an execution identifier to each command: the common prefix contains the execution identifiers in the old document, any commands that are new, or succeeding a new command are assigned a freshly generated identifier.

These pairs are then reported back to the prover, using a similar message syntax as used for the update message. The message formatting is slightly different: instead of sending function and argument as a single XML message, the header and body are separated in two messages.

\begin{lstlisting}
<protocol function="assign_update" />
<:>
  <:>
    <:>-1</:>
    <:><:>49</:></:>
  </:>
  !\vdots!
</:>
\end{lstlisting}

This assigns the execution identifier 49 to the command identifier -1. For Isabelle, it is possible to have multiple execution identifiers, so the second element of the pair is actually a list.

The PIDE plugin does nothing with this message, except storing the assignments in a lookup table. This is message is also used as a confirmation that the update was received.

\paragraph{Computation}
After the update assignment, the prover adds the commands to the STM where necessary: it finds the last common command between old and new, and instructs STM to insert new commands after this command. The insertion could not be done before sending out the assignment, because the STM parses the commands (for scheduling) and would report syntax errors before an assignment is reported: these messages would never be caught by PIDE, and not reported to the user. Any command added to the STM is accompanied by the execution identifier, that will be used by Coq to report information on that specific command.  After adding the commands, the STM is started by asking it to compute the final command in the document: this triggers the computation of all preceding commands.

During the computation, information gets reported asynchronously, including the execution identifier. This information is packed in messages and sent back to jEdit.

The first of these messages is the error message:
\begin{lstlisting}
<error serial="7" id="6">
  Error: Attempt to save an incomplete proof
</error>
\end{lstlisting}
The message has a serial number that is used internally in PIDE, and the execution identifier attached to the ``Qed.'' command. In this particular case, Coq could not recover the offending part of the command, so the message contains only the identifier. As a result, jEdit underlines the entire command. If an offset was reported, it would cause only the problematic command to be underlined.

The second is an example of hyperlinks:
\begin{lstlisting}
<report offset="9" end_offset="18" id="16">
  <entity def_id="13" id="16" offset="9" end_offset="18"
          def_offset="7" def_end_offset="16"
          name="app_assoc" kind="thm"/>
</report>
\end{lstlisting}

This message can contain multiple reports for a single execution identifier, but in this case it just describes the ``app\_assoc'' entity of kind ``thm'', Coq's internal identifier for theorems and lemmas. It includes the location information of the lemma, which in this case is within a command of the same document (it has an assigned ``def\_id'') and a certain offset within that command. The message also allows for defining the location in other files, by specifying the ``def\_file'' attribute with a line and character offset.

The final message reports goal states. The message for the displayed goal state is:
\begin{lstlisting}
<writeln serial="144" id="16">1 subgoal
n : nat
l1 : natlist
!\vdots!
============================
 app (!\ldots</writeln>
\end{lstlisting}

When receiving this message, PIDE stores the data, and retrieves it when the cursor is on the specified line.

There are other messages in the protocol, for example to report prover progress, but these three types are the ones currently implemented for Coq.

\section{Conclusions}

% Usable editing environment within three months: includes features unavailable on other platforms (Hyperlinking)
% Useful for considering the generics of prover-interaction.
Based on the original PIDE implementation for Isabelle with some minor generalizations, and by adding the protocol functions to Coq (at the time of writing, all PIDE functionality is contained in approximately 750 lines of fairly verbose OCaml code), we have obtained a functional, asynchronous editing environment for Coq. At the time of writing, this editing environment is available from a dedicated download page: \href{http://pages.saclay.inria.fr/carst.tankink/jedit.html}{http://pages.saclay.inria.fr/carst.tankink/jedit.html}.
The integration was developed in a very short period: going from the STM interface in Coq, and the existing Isabelle code base, jEdit could work with Coq in two man-months of work. This time included reading the PIDE code base in order to understand how the protocol works, and what parts are essential.

Going forwards, there still is more work to be done in making the PIDE architecture more accessible to other provers, but adding Coq is a first step in identifying what elements of the architecture are specific to a particular prover and what parts can be reused regardless of the proof assistant in use. For now, a prover that wants to benefit from the PIDE architecture, needs to supply:

\begin{itemize}
  \item a parser that can recognize command spans. This parser needs to be added to the Scala code of the PIDE architecture;
  \item a protocol implementation that can unpack the PIDE messages and provide them to the prover. This implementation is also responsible for supplying feedback and output to the interface.
\end{itemize}

These recommendations are not fixed edicts, but will provide a running system with a minimum of effort, and are therefore a good starting point for adding specialized messages: this is where we started with Coq, and we are now looking forward to a more asynchronous, declarative future of interaction.

\bibliographystyle{eptcs}
\bibliography{uitp}
\end{document}